\documentstyle[pra,aps]{revtex}

\title{On the Complexity of Quantum Searching Using Complex Queries}

\author{Markus~Grassl and Thomas~Beth}
\address{
	Institut f{\"u}r Algorithmen und Kognitive Systeme,
        Universit{\"a}t Karlsruhe, Am Fasanengarten 5, 
	D--76\,128 Karlsruhe, Germany\\
        e--mail: {\tt grassl@ira.uka.de}
}
\date{23.06.1997}

\def\ket#1{\left|#1\right>}

\def\bm#1{\mathchoice{\mbox{\boldmath{$\displaystyle #1$}}}%
{\mbox{\boldmath{$\textstyle #1$}}}%
{\mbox{\boldmath{$\scriptstyle #1$}}}%
{\mbox{\boldmath{$\scriptscriptstyle #1$}}}}
\let\ds=\displaystyle

\begin{document}

\maketitle
\begin{abstract}%
We discuss the quantum search algorithm using complex queries that has
recently been published by Grover \cite{ComplexQueries}. We recall the
algorithm adding some details showing which complex query has to be
evaluated. Based on this version of the algorithm we discuss its
complexity.
\end{abstract}

\section{Introduction}
We assume that the reader is familiar with the paper {\em Complex
Quantum Queries/Quantum computers can search arbitrarily large
databases by a single query} \cite{ComplexQueries}.

\subsection{Statement of the Problem}
The algorithm of \cite{ComplexQueries} solves the following problem:

\begin{em}
{\bf Problem:} Given a set ${\cal M}$ of $N$ items and a Boolean
function $f\colon {\cal M} \rightarrow \{0,1\}$, find an element
$x\in{\cal M}$ with $f(x)=1$ using the function
\begin{equation}\label{complexQuery}
\tilde{f}({\cal T})
  =\left|\bigl\{\bm{x}\in{\cal T}|f(x)=1\bigr\}\right|\bmod 2.
\end{equation}
where ${\cal T}\subseteq{\cal M}$ is an arbitrary subset of ${\cal
M}$. We assume w.\,l.\,o.\,g.{} $N=2^\nu$ and ${\cal
M}=\{1,2,\ldots,N\}$.
\end{em}

Grover considers $f(\bm{x})$ as an {\em elementary query} since only
one item of ${\cal M}$ is involved, whereas the {\em complex query}
$\tilde{f}({\cal T})$ depends on an arbitrary subset of ${\cal
M}$. The function checks whether the number of elements of the subset
${\cal T}$ satisfying the predicate $f$ is odd.

Another complex query is $\hat{f}({\cal T})=\exists x\in{\cal T}\colon
f(x)=1$ which can be used for binary searching.

\subsection{Definitions}
To formulate the quantum search algorithm it is helpful to consider
the following auxiliary functions.

For $j=1,\ldots,N$, the function 
\begin{equation}\label{chiFunction}
\begin{array}{rrcl}
\chi_j: & \{0,1\}^{\nu\eta}&\rightarrow&\{0,1\}\\
        & (\bm{x}_1,\ldots,\bm{x}_\eta) & \mapsto & 
    \left|\bigl\{i: i \in \{1,\ldots,\eta\}
               |\bm{x}_i=j\bigr\}\right|\bmod 2
\end{array}
\end{equation}
checks the parity of the number of $\nu$--bit--strings $\bm{x}_i$ equal
to $j$ in the $\eta$--tuple $(\bm{x}_1,\ldots,\bm{x}_\eta)$.

Furthermore, the subset ${\cal T}$ used in the complex query
$\tilde{f}({\cal T})$ is encoded by its incidence vector
$$
{\cal X}=\left(\chi_{\cal T}(1),\ldots,\chi_{\cal T}(N)\right)
$$
where $\chi_{\cal T}$ is the characteristic function of the subset
${\cal T}$. Thus, the complex query can be considered as a Boolean
function with $N$ inputs.

\section{Quantum Search Algorithm Using Complex Queries}\label{algorithm}
In this section we restate the algorithm of \cite{ComplexQueries} and
include some details showing how the necessary operations might be
implemented. 

Be  $\eta$ a constant of order $N(\log N)^2$.
\begin{enumerate}
\item Prepare the following state on $\nu\eta+N+1$ qubits:
$$
  \ket{\psi_1}=
     \underbrace{\left(\ket{0}\ldots\ket{0}\right)}_{\nu\eta}
   \otimes
     \underbrace{\left(\ket{0}\ldots\ket{0}\right)}_{N}
   \otimes\ket{0}
$$
\item 
\begin{enumerate}
  \item 
  Perform a Hadamard transform on the first $\nu\eta$ qubits and the
  last qubit, i.\,e., 
  $$
      H = \frac{1}{\sqrt{2^{\nu\eta}}}
          \left(\begin{array}{rr}
                  1 &  1 \\
                  1 & -1 
		\end{array}\right)^{\otimes \nu\eta}
          \otimes\openone_{2^{N}}
          \otimes
          \frac{1}{\sqrt{2}}
          \left(\begin{array}{rr}
                  1 &  1 \\
                  1 & -1 
		\end{array}\right)
  $$
  \item Perform a $\sigma_z$ rotation on the last qubit
\end{enumerate}
This results in the state (to simplify the notation, normalization
factors are omitted here and in the remainder of the paper):
\begin{eqnarray*} 
\ket{\psi_2}
 &=& \left(\sum_{\bm{x}=1}^N\ket{\bm{x}}\right)^{\otimes\eta}
       \otimes\left(\ket{0}\right)^{\otimes N}
       \otimes\left(\ket{0}-\ket{1}\right)\\ 
 &=& \sum_{(\bm{x}_1,\ldots,\bm{x}_\eta)\in\{1,\ldots,N\}^\eta}
               \ket{\bm{x}_1}\otimes\ldots\otimes\ket{\bm{x}_\eta}
       \otimes\underbrace{\left(\ket{0}\ldots\ket{0}\right)}_{N}
       \otimes\left(\ket{0}-\ket{1}\right).  
\end{eqnarray*}
\item For $j=1,\ldots,\eta$ add the value of the function
$\chi_j(\bm{x}_1,\ldots,\bm{x}_\eta) = \left|\bigl\{i: i \in
\{1,\ldots,\eta\} |\bm{x}_i=j\bigr\}\right|\bmod 2$ to qubit $\nu\eta+j$
resulting in the state
$$
 \ket{\psi_3}=
   \sum_{(\bm{x}_1,\ldots,\bm{x}_\eta)\in\{1,\ldots,N\}^\eta}
        \ket{\bm{x}_1}\otimes\ldots\otimes\ket{\bm{x}_\eta}
       \otimes\ket{\chi_1}\otimes\ldots\otimes\ket{\chi_N}
       \otimes\left(\ket{0}-\ket{1}\right).
$$
\item Add the value of the function $\tilde{f}({\cal X})$ where ${\cal
X}\subseteq{\cal M}$ is given by the support of the incidence vector
$(\chi_1,\ldots,\chi_N)$ to the last qubit:
\begin{eqnarray*}
 \ket{\psi_4} &=&
  \sum_{(\bm{x}_1,\ldots,\bm{x}_\eta)\in\{1,\ldots,N\}^\eta}
        \ket{\bm{x}_1}\otimes\ldots\otimes\ket{\bm{x}_\eta}
       \otimes\ket{\chi_1}\otimes\ldots\otimes\ket{\chi_N}
       \otimes\left(|0+\tilde{f}({\cal X})\rangle
                      -|1+\tilde{f}({\cal X})\rangle\right)\\
&=& \sum_{(\bm{x}_1,\ldots,\bm{x}_\eta)\in\{1,\ldots,N\}^\eta}
        \ket{\bm{x}_1}\otimes\ldots\otimes\ket{\bm{x}_\eta}
       \otimes\ket{\chi_1}\otimes\ldots\otimes\ket{\chi_N}
       \otimes(-1)^{\tilde{f}({\cal X})}\left(\ket{0}-\ket{1}\right)
\end{eqnarray*}
\item Repeat step 3 to dis--entangle the states, i.\,e., for
$j=1,\ldots,\eta$ add the value of function
$\chi_j(\bm{x}_1,\ldots,\bm{x}_\eta)$ to qubit $\nu\eta+j$ resulting in
the state
\begin{eqnarray}
\ket{\psi_5}
  &=&\sum_{(\bm{x}_1,\ldots,\bm{x}_\eta)\in\{1,\ldots,N\}^\eta}\nonumber
        \ket{\bm{x}_1}\otimes\ldots\otimes\ket{\bm{x}_\eta}
  	 \otimes\left(\ket{0}\right)^{\otimes N}
       \otimes(-1)^{\tilde{f}({\cal X})}\left(\ket{0}-\ket{1}\right)\\
&=&\left(\sum_{\bm{x}=1}^N(-1)^{f(\bm{x})}\ket{\bm{x}}\right)^{\otimes\eta}
      \otimes\left(\ket{0}\right)^{\otimes N}
      \otimes\left(\ket{0}-\ket{1}\right).\label{idStep5}
\end{eqnarray}
(As equality (\ref{idStep5}) is not obvious, it is proved separately
in section \ref{proofStep5}.)
\item Apply the operator $D$ ({\em inversion about average})
\begin{equation}\label{invAverage}
      D = \frac{1}{\sqrt{2^{n}}}
          \left(\begin{array}{rr}
                  1 &  1 \\
                  1 & -1 
		\end{array}\right)^{\otimes n}
    \cdot
          \left(\begin{array}{rrrr}
                  -1 &   \\
                    & 1 \\
                    & & \ddots \\
                    & & & 1
		\end{array}\right)^{\otimes n}
    \cdot
          \frac{1}{\sqrt{2^{n}}}
          \left(\begin{array}{rr}
                  1 &  1 \\
                  1 & -1 
		\end{array}\right)^{\otimes n}
\end{equation}
on each of the first $\eta$ registers of length $n$. After this step,
the system is in a state that consists of $\eta$ (non--entangled)
copies of the state after one step of the original quantum search
algorithm \cite{bounds,searchSTOC,searchPRL}:
\begin{equation}\label{oneSearchStep}
\ket{\psi_6}=
\left(\sum_{f(\bm{x})=1}k_{\bm{x}}\ket{\bm{x}}
      +\sum_{f(\bm{x})=0}l_{\bm{x}}\ket{\bm{x}}\right)^{\otimes\eta}
      \otimes\left(\ket{0}\right)^{\otimes N}
      \otimes\left(\ket{0}-\ket{1}\right)
\end{equation}
with suitable values $k_{\bm{x}}$ and $l_{\bm{x}}$ as to be described
in section \ref{amplitudes}.
\item By measuring each of the first $\eta$ registers of length $n$, a
set ${\cal S}=\{\tilde{\bm{x}}_1,\ldots,\tilde{\bm{x}}_\eta\}$ of
$\eta$ samples is obtained.
\item In a (classical) post--processing step, an element $\bm{x}_0$ of
the $\eta$ samples $\tilde{\bm{x}}_i$ with maximal frequency is
searched. This element $\bm{x}_0$ is the output of the algorithm.
\end{enumerate}

\section{Discussion}
\subsection{Proof of the Identity in Step 5}\label{proofStep5}
From the definition (\ref{complexQuery}) of $\tilde{f}({\cal X})$ and
the definition (\ref{chiFunction}) of $\chi_j$ it follows that 
\begin{equation}\label{identity}
\tilde{f}({\cal X})=\tilde{f}\left(\chi_1,\ldots,\chi_N\right)=
\sum_{i=1}^{\eta}f(\bm{x}_i)\bmod 2.
\end{equation}
This identity can be proved by the so--called method of double
counting. For doing so note that each of the functions $\chi_j$ and
thus ${\cal X}$ depend on
$\underline{\bm{x}}=(\bm{x}_1,\ldots,\bm{x}_\eta)$. Thus,
$$
\begin{array}{rclcl}
\tilde{f}({\cal X}(\underline{\bm{x}}))
 &=& \tilde{f}({\rm  support}
        (\chi_1(\underline{\bm{x}}),\ldots,\chi_N(\underline{\bm{x}})))
 &=& |\{j\in\{1,\ldots,N\}|
        \chi_j(\underline{\bm{x}})=1\wedge f(j)=1\}| \bmod 2\\
 &=& \ds\sum_{j=1}^N f(j)\chi_j(\underline{\bm{x}}) \bmod 2
 &=& \ds\sum_{j=1}^N f(j)\sum_{i=1}^\eta\delta_{\bm{x}_i,j} \bmod 2\\
 &=& \ds\sum_{i=1}^\eta\sum_{j=1}^N \delta_{\bm{x}_i,j}f(j) \bmod 2
 &=& \ds\sum_{i=1}^\eta f(\bm{x}_i) \bmod 2.
\end{array}
$$
Equation (\ref{identity}) implies
$$
(-1)^{\tilde{f}({\cal X(\underline{\bm{x}})})}
  \ket{\bm{x}_1}\otimes\ldots\otimes\ket{\bm{x}_\eta}
=\left((-1)^{f(\bm{x}_1)}\ket{\bm{x}_1}\right)
    \otimes\ldots\otimes
 \left((-1)^{f(\bm{x}_\eta)}\ket{\bm{x}_\eta}\right)
$$
which proves the identity (\ref{idStep5}).
\subsection{The Probability of Success}\label{amplitudes}
Be $t$ the number of elements of ${\cal M}$ that satisfy the predicate
$f$, Then, the amplitudes $k_{\bm{x}}$ and $l_{\bm{x}}$ in
(\ref{oneSearchStep}) are
$$
k_{\bm{x}}=\left(3-\frac{4t}{N}\right)\frac{1}{\sqrt{N}}
\qquad\mbox{and}\qquad
l_{\bm{x}}=\left(1-\frac{4t}{N}\right)\frac{1}{\sqrt{N}}.
$$
Thus, the probabilities to measure an element $\tilde{\bm{x}}_i$ with
$f(\tilde{\bm{x}}_i)=1$ (or with $f(\tilde{\bm{x}}_i)=0$) are
\begin{eqnarray}
Pr[\tilde{\bm{x}}_i|f(\tilde{\bm{x}}_i)=1]\label{probf1}
  &=& \left(9-\frac{24t}{N}+\left(\frac{4t}{N}\right)^2\right)\frac{1}{N}\\
\noalign{and}
Pr[\tilde{\bm{x}}_i|f(\tilde{\bm{x}}_i)=0]\label{probf0}
  &=& \left(1-\frac{8t}{N}+\left(\frac{4t}{N}\right)^2\right)\frac{1}{N}.
\end{eqnarray}
The output of the algorithm is an element $\bm{x}_0$ with maximal
frequency in the sample ${\cal S}$ of size $\eta$. Using
(\ref{probf1}) and (\ref{probf0}), the probability that $\bm{x}_0$
satisfies $f$, i.\,e., $Pr[f(\bm{x}_0)=1]$, might be calculated
exactly given the values $N$, $t$, and $\eta$. 

Using the central limit theorem, in \cite{ComplexQueries} it is shown
that $Pr[f(\bm{x}_0)=1]$ approaches one for $\eta$ of order
$N\left(\log N\right)^2$.

\subsection{The Complexity of the Algorithm}
In the following we consider the complexity of the steps of the
algorithm as presented in section \ref{algorithm}. The number of
elementary (two--bit) gates (cf.~\cite{gates}) will be used as a
measure for the complexity of the quantum operations.
\begin{itemize}
\item the number of Hadamard transforms:\\
For the preparation of the equal superposition in step 1, $\nu\eta$
Hadamard transforms are needed. Twice that number is needed to apply
the operator $D^{\otimes\eta}\otimes\openone$ for the inversion about
average on the first $\nu\eta$ qubits.

\item computation of $\chi_j$ in steps 3 and 5:\\ 
The function $\chi_j$ can be computed in the following manner: for
$i=1,\ldots,N$ add the value of the function
$\delta_j(\bm{x}_i)=\delta_{\bm{x}_i,j}$ to qubit $\nu\eta+j$. In the
language of \cite{gates}, these are $\sigma_x$ rotations conditioned
on $\nu$ qubits ($\bigwedge_\nu(\sigma_x)$) which can be achieved with
${\cal O}(\nu)$ elementary operations.

This gives a total complexity of ${\cal O}(\nu\eta)$ for steps 3 and 5.

Note that even though after step 5 the second register of $N$ qubits
has been reset to $(\ket{0})^{\otimes N}$, it was needed for the
superimposed computation of $\tilde{f}({\cal X})$ which indeed is a
function $tilde{f}({\cal X}(\bm{x}_1,\ldots,\bm{x}_\eta))$. 
\item inversion about average:\\ 
To compute the operator $D^{\otimes\eta}\otimes\openone$ (inversion
about average) besides the Hadamard transforms the conditional phase
change (the diagonal matrix in (\ref{invAverage})) has to be applied
$\eta$ times. These operations are $\sigma_z$ rotations conditioned on
$\nu$ qubits ($\bigwedge_\nu(\sigma_z)$) and can be achieved with totally
${\cal O}(\nu\eta)$ elementary operations.

\item classical post--processing:\\ 
To find an element $\bm{x}_0$ of maximal frequency in the sample
${\cal S}$ the sequence of samples has to be sorted which has
complexity ${\cal O}(\eta\log\eta)$.
\end{itemize}
The complexity of the algorithm is dominated by the complexity of
steps 3 and 5, possibly the evaluation of $\tilde{f}({\cal X})$ in
step 4, and of the (classical) post--processing in step 8. In summary,
the complexity of the quantum search algorithm using complex queries
is ${\cal O}(\nu\eta)+{\cal O}(\eta\log\eta)$ plus the complexity of one
evaluation of the complex query $\tilde{f}$. For $\eta$ of order
$N(\log N)^2$, the algorithm has a complexity of ${\cal
O}\left(N^2(\log N)^2\right)$.

\section{Conclusions}
The main result of \cite{ComplexQueries} is that the complex query
$\tilde{f}({\cal T})$ has to be evaluated only once instead of ${\cal
O}(\sqrt{N/t})$ evaluations of the elementary query $f(\bm{x})$ in
\cite{searchSTOC}. Compared to the original quantum search algorithm
\cite{searchSTOC}, both the number of qubits and the number of quantum
and classical operations to be performed are dramatically
increased. Thus there is a trade--off between the number of elementary
operations and the complexity of the complex query.

It has to be investigated in which situations it might be easier to
evaluate the complex query once than to evaluate the elementary query
many times.

Nevertheless, the quantum search algorithm \cite{ComplexQueries}
proves that with a quantum computer only one query evaluation is
needed whereas any classical algorithm will be limited by the
information theoretic bound of at least logarithmic many queries.

\section{Acknowledgements}
The authors would like to thank Lov Grover for his suggestion to make
this intentionally private comments public.

\end{document}